\documentclass[prl,aps,twocolumn,floatfix,showpacs]{revtex4}

\usepackage{graphicx}
\usepackage{amsmath}
\usepackage{comment}
\usepackage{braket}
\usepackage{yfonts}
\usepackage{MnSymbol}
\newcommand{\be}{\begin{equation}}
\newcommand{\ee}{\end{equation}}
\newcommand{\ba}{\begin{aligned}}
\newcommand{\ea}{\end{aligned}}

\newcommand{\gothb}{\text{\textgoth{b}}}
\newcommand{\gothc}{\text{\textgoth{c}}}

\begin{document}
\title{\centering Dynamical Phase Transitions as Properties of the Stationary State:\\Analytic Results after Quantum Quenches in the Spin-1/2 XXZ Chain} 
\author{Maurizio Fagotti}
\affiliation{\mbox{The Rudolf Peierls Centre for Theoretical Physics,
    Oxford University, Oxford, OX1 3NP, United Kingdom}}
\begin{abstract}
The (Loschmidt) overlap between the state at different times after a quantum quench is attracting increasing interest, as it was recently shown that in the thermodynamic limit  its logarithm per unit of length has a non-analytic behavior if a Hamiltonian parameter is quenched across a critical point. 
This phenomenon was called a ``dynamical phase transition'' in analogy with the behavior of the canonical partition function at an equilibrium phase transition. We distinguish between local and nonlocal contributions to the aforementioned quantity and derive an analytic expression for the time evolution of the local part after quantum quenches in the XXZ \mbox{spin-$\frac{1}{2}$} chain.
The state that describes the stationary properties of (local) observables can be represented by a Gibbs ensemble of a generalized Hamiltonian; 
we reveal a deep connection between the appearance of singularities and the excitation energies of the generalized Hamiltonian.
\end{abstract}

\maketitle %

\paragraph{Introduction.} %

It is striking that after a century one of the fundamental goals of quantum mechanics, understanding the time evolution of a state in a closed system, is still a focus of debate. In the last ten years, motivated by the extraordinary experimental advances in the realization of highly tunable, practically closed quantum systems~\cite{quench_exp},  theoretical investigations have revealed novel connections between the non-equilibrium time evolution of observables and the properties of the Hamiltonian~\cite{quenchrev}. 

We focus on the simplest nonequilibrium situation of a (global) quantum quench. 
The system is originally prepared in the ground state $\ket{\Psi_0}$ of  a translational invariant Hamiltonian $H(g_0)$ with short range interactions, where $g_0$ is an experimentally tunable parameter. Then the parameter is suddenly changed to a different value $g$ in such a way that the state unitarily evolves with the new Hamiltonian $H(g)$: $\ket{\Psi(t)}=e^{-i H(g) t}\ket{\Psi_0}$. 

In a quantum many-body system, $\ket{\Psi(t)}$ contains an incredibly large amount of information that is impossible to manage. 
One of the reasons why quantum quenches have been attracting much attention is however that generally the dynamics of local degrees of freedom admits a simplified description. In particular, at late times after the quench, at the subsystem level, the state can be replaced by a statistical ensemble, which is completely characterized by the expectation values of the local conservation laws~\cite{gge,initstate,quenchdephase,RDM, TBAgge,ggeXXZ,ggeXXZ2,gge1,fluctgge,Cazathermal,MK:2009,TFIC0, KCC:2013}. 
For quenches in generic models, this generally results in a thermal ensemble with an effective temperature fixed by energy conservation~\cite{abstherm}.
For quenches in integrable models, in which instead there are an infinite number of local conservation laws, the stationary state is generally described by the so-called generalized Gibbs ensemble (GGE)~\cite{gge}
\be\label{eq:rhoGGE}
\rho_{\rm GGE}=\frac{1}{Z}e^{-\sum_{i=1}\lambda_{i} H^{(i)}}\qquad [H^{(i)},H^{(j)}]=0\, ,
\ee
where $H^{(i)}$ are the charges with local density~\cite{RDM} and $H^{(1)}$ is the Hamiltonian.

Statistical descriptions also succeed in explaining some dynamical properties. 
For example, semiclassical theories~\cite{quenchMus,semiclas,quenchCFT} have proven capable of capturing qualitative aspects of the time evolution; in addition, the asymptotic relaxation to a stationary state is a crucial aspect of the framework, proposed in  Ref.~\cite{CE:2013}, to analytically compute the large time behavior of observables. 
Therefore, in spite of the complexity of the problem, the main aspects of the time evolution of correlation functions~\cite{quenchCFT,TFIC0,numtimecorr,quenchXXZ,quenchint,quenchMus,quenchFT,quenchLutt,Luttuniv,CE:2013,semiclas} and entanglement entropies~\cite{quenchent}, as well as some universal features of statistical fluctuations~\cite{work} and the response of the system to small perturbations~\cite{TFICd}, have been understood. 

In a recent work~\cite{dpt0} on global quenches across the critical point in the transverse-field Ising chain, attention was drawn to the appearance of non-analytic behavior in
\be\label{eq:dfen}
f(t)=-\lim_{L\rightarrow\infty}\frac{1}{L}\log G(t)\, ,
\ee
where $L$ is the system size and $G(t)$ is the Loschmidt overlap
\be\label{eq:GL}
G(t)=\braket{\Psi(t_0)|\Psi(t+t_0)}=\braket{\Psi_0|e^{-i Ht}|\Psi_0}\, .
\ee
By extending $G(t)$ to complex time and interpreting it as a boundary partition function in the complex plane, it was shown that singularities
 arise from the zeros of the partition function that, in the thermodynamic limit, coalesce to lines that cut the real axis of time.  This was reminiscent of  the equilibrium case in which the zeros of the partition function in the complex plane of inverse temperature approach the real axis exactly at the critical temperature. For that reason Heyl \emph{et al}~\cite{dpt0} coined the term ``dynamical phase transition''. 
Analogously, we will refer to $f(t)$ as the ``dynamical free energy density''.

Ref.~\cite{dpt1} provided numerical evidence that this picture can be extended to interacting models (also with integrability breaking terms). 

However, an adequate understanding of the phenomenon is still lacking, and one of the reasons is that, so far, the theoretical analysis was limited to noninteracting models. 
 
Since we are ultimately interested in possible effects on the time evolution of local observables, we analyze the ``bulk part'' of the dynamical free energy density~\eqref{eq:dfen}.
We point out the latter is a property of the stationary state after the quench, \emph{i.e.}, in integrable models, of the generalized Gibbs ensemble~\eqref{eq:rhoGGE}. 
Using the formalism developed in Ref.~\cite{ggeXXZ}, we investigate quenches in the antiferromagnetic spin-1/2  XXZ chain.
In noninteracting models the non-analyticities of the overlap \eqref{eq:GL} can be associated with the zeros of the dispersion relation of the \emph{generalized Hamiltonian} (\emph{cf}. Eq.~\eqref{eq:rhoGGE})
\be\label{eq:genH}
H_{\rm GGE}=\sum_i\lambda_i H^{(i)}\, .
\ee
We obtain a similar result for the bulk part of $f(t)$ after quenches in the (interacting) XXZ chain.
We therefore reinterpret the non-analyticities as the effect of the absence of a gap in the excitation energy of $ H_{\rm GGE}$. 

\paragraph{Loschmidt amplitude and generalized Gibbs ensemble.}%

After a global quench the energy (above the ground state) is extensively high and consequently the overlap $G(t)$~\eqref{eq:GL} is exponentially small in the system size. This accounts for the definition~\eqref{eq:dfen} of the dynamical free energy density $f(t)$.
Since $G(t)$ is the expectation value of an operator that commutes with the Hamiltonian, it is determined only by the time independent elements of the density matrix $\ket{\Psi(t)}\bra{\Psi(t)}$, which can be formally represented by the diagonal ensemble \mbox{$\rho_{\rm DE}\sim  \lim_{t\rightarrow\infty}\frac{1}{t}\int_{0}^t \ket{\Psi(\tau)}\bra{\Psi(\tau)}\mathrm d\tau$}.
Thus we have
\be\label{eq:f}
f(t)=-\lim_{L\rightarrow\infty}\frac{1}{L}\log \mathrm{Tr}[\rho_{\rm DE} e^{-i H t}]\, .
\ee
This can be formally written as a power series whose coefficients are the energy cumulants $c_n$ per unit of length, \emph{i.e.} \mbox{$f(t)=-\sum_{n=1}c_n(-i t)^n/n!$}~\cite{work}. Here, \emph{e.g.}, $c_1$ is the energy density and $c_2$ is the squared energy fluctuation per unit of length.

The cumulant expansion is useful to distinguish local (bulk) contributions from nonlocal ones. 
In order to clarify this point we work out the first two cumulants. By writing the Hamiltonian as $H=\sum_\ell \mathcal H_\ell$, where $\mathcal H_\ell$ is the (local) energy density operator, we have $c_1=\mathrm{Tr}[\rho_{\rm DE} \mathcal H_1]$.
By contruction, the expectation value of $\mathcal H_1$ can be computed in  the GGE, \emph{i.e.} $c_1=\mathrm{Tr}[\rho_{\rm GGE} \mathcal H_1]$. 
The second cumulant $c_2$ is the sum of the connected two-point functions of the energy density operator
\be\label{eq:c2}
c_2=\lim_{L\rightarrow\infty}\sum_{\ell=1-\frac{L}{2}}^{\frac{L}{2}}\mathrm{Tr}[\rho_{\rm DE} \mathcal H_1\mathcal H_{1+\ell}]-\mathrm{Tr}[\rho_{\rm DE} \mathcal H_1]^2\, .
\ee
We identify the bulk part of $c_2$ as the sum of connected correlations with distance $\ell\ll L\rightarrow\infty$. This involves operators well-defined in the thermodynamic limit. 
Since $\rho_{\rm GGE}$ is locally equivalent to $\rho_{\rm DE}$, such expectation values can be computed in the generalized Gibbs ensemble~\cite{TFIC0,RDM,mine2,initstate}. 
The remaining contributions to \eqref{eq:c2} involve nonlocal operators  that stretch along the entire chain.

An analogous discussion holds true for the higher order cumulants, where we can still single out the bulk contribution, described by the GGE, from the rest, which is the sum of connected $n$-point functions in which at least one distance scales with the system size. 
Having this picture in mind, we identify the bulk part of $f(t)$ as 
\be\label{eq:bf}
f_{\rm bulk}(t)=-\lim_{L\rightarrow\infty}\frac{1}{L}\log \mathrm{Tr}[\rho_{\rm GGE} e^{-i H t}]\, .
\ee
We illustrate the importance of bulk contributions with a quench in the transverse-field Ising chain.

\paragraph{Quantum Ising model}%
The Hamiltonian of the transverse field Ising chain (TFIC) can be written as
\be
H_{\rm I}^{(h)}=-\frac{1}{2}\sum_{\ell=1}^L\Bigl[\sigma_\ell^x \sigma_{\ell+1}^x+h \sigma_\ell^z\Bigr]\, ,
\ee
where $\sigma_\ell^{\alpha}$ are Pauli matrices, $\sigma^\alpha_{L+1}\equiv \sigma_1^\alpha$, and we assume $h>0$.  
The phase diagram is characterized by a critical point at $h=1$ that separates a ferromagnetic phase \mbox{$(h<1)$} from a paramagnetic one \mbox{$(h>1)$}.
The model is exactly solvable: it is mapped to noninteracting fermions by a Jordan-Wigner transformation and then diagonalized by a Bogolioubov transformation in Fourier space.

The dynamical free energy density is given by~\cite{work,dpt0}
\be\label{eq:fIsing}
f(t)=i E_0t-\int_0^\pi\frac{\mathrm d k}{2\pi}\log\Bigl[\cos^2\frac{\Delta_k}{2}+e^{-2i \varepsilon_k t}\sin^2\frac{\Delta_k}{2}\Bigr]\, ,
\ee
where $E_0$ is the ground state energy, $\Delta_k$ is the difference between the Bogolioubov angles of the Hamiltonian after and before the quench and \mbox{$\varepsilon_k=\sqrt{1+h^2-2h\cos k}$} is the dispersion relation (see \emph{e.g.} \cite{TFIC0} for further details).
Because the initial state is eigenstate of an infinite number of conservation laws, the GGE is globally different from the diagonal ensemble~\cite{mine2}.
The GGE can be represented as~\cite{TFIC0}
\be\label{eq:GGEIs}
\rho_{\rm GGE}=
\Bigl(\lim_{L\rightarrow\infty}\Bigr)e^{-\sum_k\varepsilon_{h;h_0}^{\rm GGE}(k)(b^\dag_k b_k-1/2)+\log\frac{|\sin\Delta_k|}{2}}\, ,
\ee
where $\varepsilon_{h;h_0}^{\rm GGE}(k)= 2\mathrm{arctanh}(\cos\Delta_k)$ is the dispersion relation of the generalized Hamiltonian~\eqref{eq:genH}. One can easily show 
\be\label{eq:fbIsing}
f_{\rm bulk}(t)=i E_0t-\int_0^\pi\frac{\mathrm d k}{\pi}\log \Bigl[\cos^2\frac{\Delta_k}{2}+e^{-i \varepsilon_k t}\sin^2\frac{\Delta_k}{2}\Bigr]\, .
\ee
There is a simple relation between \eqref{eq:fbIsing} and \eqref{eq:fIsing}, \mbox{$f(t)=f_{\rm bulk}(2 t)/2$}, and in both cases singularities  appear after quenches across the critical point, \emph{i.e.} when there are momenta such that $\cos\Delta_k=0$.  
Importantly, \emph{these are gapless modes of the generalized Hamiltonian}.

Apparently $f_{\rm bulk}(t)$ provides the same information as $f(t)$. 
However the appearance of singularities in $f_{\rm bulk}(t)$, which, by definition, is determined only by local operators, is a strong indication we can establish some connection with the time evolution of local observables. And indeed the oscillation frequency of the order parameter one-point function $\braket{\Psi_t|\sigma^x_1|\Psi_t}$~\cite{TFIC0} after a quench across the critical point from the ferromagnetic phase is equal to the frequency of singularities in $f_{\rm bulk}(t)$ (\emph{cf.} \cite{dpt0,dpt1}). 

The relation with local degrees of freedom can be better understood by considering a quench from the ground state of a TFIC in which the magnetic field has a smooth global dependence on the position, $h_\ell=h_0+\delta h_0 \cos(2\pi \ell/L)$, with $0<h_0-\delta h_0<1<h_0+\delta h_0$.
In order to properly define a thermodynamic limit we must now specify the ``global position'' $x_0$, such that $\ell/L\sim x_0$. The initial state is then locally equivalent to the ground state of the TFIC with magnetic field $h(x_0)\equiv h_0+\delta h_0\cos(x_0)$. The bulk part of $f(t)$ is again given by Eq.~\eqref{eq:fbIsing}. The function is non-analytic whenever $h(x_0)$ and the final magnetic field correspond to different phases; importantly, the relation with the oscillation frequency of $\braket{\Psi_t|\sigma^x_\ell|\Psi_t}$ (with $\ell/L\sim x_0$) still holds.
On the other hand $f(t)$ is independent of $x_0$, and hence not directly connected with the behavior of local observables.

\paragraph{XXZ model.} %

We consider the antiferromagnetic \mbox{spin-$\frac{1}{2}$} Heisenberg XXZ chain with Hamiltonian
\be
\label{eq:HXXZ}
H^{(\Delta)}_{\rm XXZ}=\frac{1}{4}\sum_{\ell=1}^L\Bigl[\sigma_\ell^x\sigma_{\ell+1}^x+\sigma_\ell^y\sigma_{\ell+1}^y+\Delta (\sigma_\ell^z\sigma_{\ell+1}^z-1)\Bigr]\, ,
\ee
where $\Delta$ is the anisotropy parameter and $\sigma_{L+1}^\alpha\equiv \sigma_1^\alpha$. The Hamiltonian is gapless for $|\Delta|\leq 1$ and for $\Delta>1$ the ground state is antiferromagnetic. The model is solvable by the algebraic Bethe Ansatz method~\cite{invscat}, which gives the local conservation laws as the logarithmic derivative of the transfer matrix $\tau$ at the shift point (see \emph{e.g.} Ref.~\cite{ggeXXZ} for the exact definition): 
\be\label{eq:LCL}
H^{(k)}=i\Bigl(\frac{\sinh \eta}{\eta}\frac{\partial}{\partial \lambda}\Bigr)^k\log\tau(i+\lambda)\Bigr|_{\lambda=0}\, .
\ee
Here $\eta$ parametrizes the anisotropy \mbox{$\Delta=\cosh\eta$}. 
In Refs~\cite{ggeXXZ, ggeXXZ2} the formalism for computing thermal correlators in the massive phase of the XXZ chain (see \emph{e.g.} \cite{thermalcorr} and references therein) was adapted to the GGE that results from the nonequilibrium evolution $e^{-iH_{XXZ}^{(\Delta)}t}\ket{\Psi_0}$ with $\Delta>1$. In particular, Ref.~\cite{ggeXXZ} derived a system of nonlinear integral equations that takes as input the expectation value of the local conservation laws in the initial state. 
The system of equations is reported in \cite{SM}; here we show the general form of the relevant equations. 
For the sake of simplicity we assume zero longitudinal magnetization $\braket{\frac{1}{L}\sum_\ell\sigma_\ell^z}=0$ and parity invariant initial states $\braket{\frac{1}{L}H^{(2n)}}=0$. The expectation values in the GGE can then be expressed in terms of a complex $\pi$-periodic function $\gothb(x)$ that satisfies 
\be\label{eq:nlib}
\mathcal E[\gothb](x)=\frac{\sinh \eta}{2}\sum_{j=0}\lambda_{2j+1}\Bigl(\frac{\sinh\eta}{2}\frac{\partial}{\partial x}\Bigr)^{2j} d(x)
\ee
where $\mathcal E$ is a nonlinear functional of $\gothb(x)$~\cite{SM} independent of the initial state and $d(x)=\sum_{n}\frac{e^{2inx}}{\cosh(\eta n)}$; $\lambda_j$ are the Lagrange multipliers of the generalized Gibbs ensemble~\eqref{eq:rhoGGE}. 
We notice that $\mathcal E[\gothb](x)$ can be interpreted as the two-spinon excitation energy of $H_{\rm GGE}$ above the ground state of $H_{\rm XXZ}^{(\Delta)}$. As long as the ground state of $H^{(\Delta)}_{\rm XXZ}$ is also  ground state of $H_{\rm GGE}$, these are low-lying excitations of the generalized Hamiltonian.  We will refer to $\mathcal E[\gothb](x)$ as the ``dressed energy'' of $H_{\rm GGE}$ (\emph{cf}. \cite{MT:2008}). 

The thermodynamic properties can be extracted from the partition function $Z$, which, for large $L$, satisfies
\be\label{eq:nliz}
\frac{\log Z}{L}\equiv \frac{\log\mathrm{Tr}[e^{-\sum_i\lambda_i H^{(i)}}]}{L}=\mathcal F[\gothb]\, ,
\ee
where $\mathcal F$ depends on the initial state only through $\gothb$. The bulk part~\eqref{eq:bf} of the dynamical free energy density reads
\be
f_{\rm bulk}(t)=\mathcal F[\gothb]-\frac{\log \mathrm{Tr}[e^{-\sum_i\lambda_i H^{(i)}-i t H^{(1)}}]}{L}\, .
\ee
The second term has the same form of \eqref{eq:nliz} but with the Lagrange multiplier of the Hamiltonian shifted by $i t$. It can then be expressed in terms of a auxiliary function $\gothc_t$ that satisfies \eqref{eq:nlib} with $\lambda_1$ replaced by $\lambda_1+it$, \emph{i.e.}
\be\label{eq:nlic}
\exp[-\mathcal E[\gothc_t](x)]=e^{-i t \frac{\sinh\eta}{2} d(x)}\exp[-\mathcal E[\gothb](x)]\, .
\ee
More explicitly, $\gothc_t$ is the solution of the integral equation
\begin{multline}\label{eq:gammat}
\gothc_t(x)=e^{-i\frac{\sinh(\eta) t}{2}d(x)}\exp\Bigl\{-\mathcal E[\gothb](x)+\Bigl[k\ast\log(1+\gothc_t)\Bigr](x)\\
-\Bigl[k_+\ast\log(1+\gothc_t)\Bigr](-x)\Bigr\}\, ,
\end{multline}
where \mbox{$k(x)=\sum_{n}\frac{e^{2i n x}}{e^{2 \eta |n|}+1}$}, \mbox{$k_\pm(x)=k(x\pm i\eta\mp i 0^+)$}, and \mbox{$[g_1\ast g_2](x)=\int_{-\pi/2}^{\pi/2}\frac{\mathrm d y}{\pi}g_1(x-y)g_2(y)$}.
The function $f_{\rm bulk}(t)$ can be finally recast in the compact form
\be\label{eq:ftexp}
f_{\rm bulk}(t)=i E_0 t-\int_{-\frac{\pi}{2}}^{\frac{\pi}{2}}\frac{\mathrm d x}{\pi}\log\Bigl[\frac{1+\gothc_t(x)}{1+ \gothb(x)}\Bigr]d(x)\, ,
\ee
where $E_0$ is the ground state energy density of $H_{XXZ}^{(\Delta)}$, \emph{i.e.}  \mbox{$E_0=-\sinh(\eta)k(0)$}. 

\begin{figure}
\includegraphics[width=0.45\textwidth]{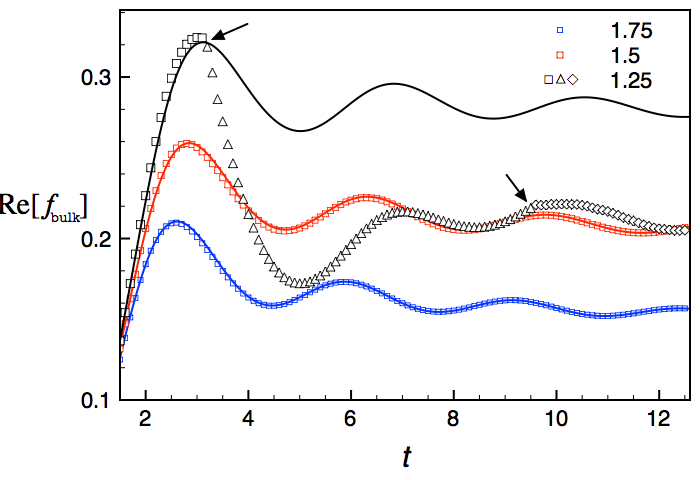}
\caption{The real part of $f_{\rm bulk}(t)$ for several interaction quenches from anisotropy $\Delta_0=+\infty$. The final anisotropy $\Delta$ is reported in the legend. The curves are the predictions based on Eq.~\eqref{eq:ansatz}. 
For $\Delta=1.25$ we find solutions for which $1+\gothc_t(x)$ has winding number equal to zero ($\medsquare$), one ($\medtriangleup$), or two ($\meddiamond$). The arrows point at the cusps.
}
\label{fig:1}
\end{figure}

Eqs~\eqref{eq:gammat} and \eqref{eq:ftexp} are well-defined only if $\gothc_t$ is sufficiently regular.
A qualitative analysis of the regular cases is however sufficient to study the emergence of non-analytic behavior:
For initial states with a finite correlation length, $1+\gothb(x)$ is generally a nonzero smooth function with zero winding number about the origin~\cite{ggeXXZ1}. If $\gothc_t(x)$ meets the same conditions, Eq.~\eqref{eq:ftexp} is smooth. If instead at the time $t^\ast$
\be\label{eq:crit}
\gothc_{t^\ast}(x_c(t^\ast))=-1
\ee
for some $x_c(t^\ast)\in(-\frac{\pi}{2},\frac{\pi}{2})$, then $f_{\rm bulk}(t)$ develops a non-analyticity at $t=t^\ast$ (as the integrands of Eqs~\eqref{eq:gammat} and \eqref{eq:ftexp} have logarithmic singularities).
After the time $t^\ast$, Eqs~\eqref{eq:gammat}\eqref{eq:ftexp} might not correctly describe the time evolution of the (bulk) dynamical free energy density; 
nevertheless, we considered some cases in which $f_{\rm  bulk}(t)$ displays a singular behavior (see \emph{e.g.} Fig.~\ref{fig:1}), resolving the ambiguities by 
 imposing continuity of $f_{\rm bulk}(t)$ (going forward in time). 
The validity of these assumptions will be investigated in a future work. 

\paragraph{Small quench.}%

Ref.~\cite{ggeXXZ} defined the limit of small quench as one for which $|\gothb(x)|\ll 1$ (and consequently $\exp(-\mathcal E[\gothb])\approx \gothb $).
Therefore, at the lowest order in $\gothb(x)$, $\gothc_t(x)\approx \gothb(x) e^{-i\frac{\sinh(\eta) t}{2}d(x)}$ (\emph{cf}. \eqref{eq:gammat}) and the (bulk) dynamical free energy density is analytic and approaches a stationary value as a power law; in particular, for a quench of the anisotropy parameter, \mbox{$\gothb(0)=\gothb(\pi/2)$=0}~\cite{ggeXXZ1} and $f_{\rm bulk}(t)$ relaxes as $t^{-3/2}$.

\paragraph{Large time.}%

In the limit of large time Eq.~\eqref{eq:gammat} can be worked out using that $\gothc_t$ is proportional to a rapidly oscillating phase. 
We propose an effective description based on the ansatz
\be\label{eq:ansatz}
\gothc_t(x)\sim e^{-i\frac{\sinh(\eta) t}{2}d(x)}e^{-\varepsilon(x)}\, .
\ee
If we restrict ourselves to nonnegative $\mathcal E[\gothb](x)$, by inserting \eqref{eq:ansatz} into \eqref{eq:gammat} and taking the time average of the convolutions, we obtain $\varepsilon(x)=\mathcal E[\gothb](x)$.
Eq.~\eqref{eq:ansatz} does not exactly describe the  large time asymptotics of $c_t(x)$, however $c_t(x)$ is only an integrated variable in \eqref{eq:ftexp} and  the ``regularized'' ansatz~\eqref{eq:ansatz} provides an excellent approximation for $f_{\rm bulk}$ when the latter is smooth (\emph{cf}. Fig.~\ref{fig:1}). 

Let us use this framework to analyze the behavior of $f_{\rm bulk}(t)$ as a function of the initial state in the neighborhood of the ground state of $H_{\rm XXZ}^{(\Delta)}$, in which $\mathcal E[\gothb](x)$ is positive. Condition \eqref{eq:crit} is satisfied (\emph{i.e.} $f_{\rm bulk}(t)$ is non-analytic) only if  $\mathcal E[\gothb](x)$ develops some zeros. 
Since $(0\leq )\mathcal E[\gothb](x)$ is the dressed energy of $H_{\rm GGE}$, this semiquantitative analysis suggests that 
\emph{singularities at large time are an effect of the relaxation to a ``thermal state'' in which the excitation energy of the generalized Hamiltonian is gapless}.

In practice our numerical analysis indicates that, as a function of the Hamiltonian parameters, singularities appear \emph{before} the dressed energy gap is closed. Fig.~\ref{fig:1} shows the real part of $f_{\rm bulk}(t)$ after the interaction quench \mbox{$\Delta_0=+\infty\rightarrow\Delta$}, with $\Delta>1$. Despite $\mathcal E[\gothb](x)>0$, for anisotropy close to the critical point we obtain a non-analytic result.
We do not have a definite explanation for this behavior. We note, however, the generalized model is at the finite (effective) temperature $T=1$ (\emph{cf.}~\eqref{eq:rhoGGE}\eqref{eq:genH}), but in the previous discussion we considered excitations at zero temperature. 
It could be worth to investigate whether finite-temperature excitations~\cite{YY} play some role; however the question is still open.

It is important to note that we have observed some non-analytic behavior after quenches within the gapped phase of the XXZ model (Fig.~\ref{fig:1}), providing evidence that \emph{the appearance of non-analyticities in $f_{\rm bulk}(t)$ is not always associated with the crossing of a critical point}. 

In contrast to the TFIC, in the XXZ model the singularities of $f_{\rm bulk}(t)$ are not periodic. However at late times $\gothc_t(x)$ depends on the time through a rapidly oscillating phase (\emph{cf}. Eq.~\eqref{eq:gammat}) which, for a generic small interval of $x$, spans the full complex unit circle. It is therefore reasonable to expect that at late times the solution $x_c$ of \eqref{eq:crit} becomes time-independent and therefore periodicity is eventually recovered. For example, assuming \eqref{eq:ansatz}, for nonnegative $\mathcal E[\gothb](x)$
the times at which $f_{\rm bulk}(t)$ is non-analytic tend to the sequence
\be\label{eq:tplus}
t^\ast(x_c)=\frac{2}{\sinh\eta}\frac{2\pi \bigl(n+\frac{1}{2}\bigr)
}{d(x_c)}\, ,
\ee
where $\mathcal E[\gothb](x_c)=0$ and $n$ is integer. 

\paragraph{Conclusions.} %

We have considered the overlap between the state at different times after a global quench. 
We singled out the contribution of local operators to the dynamical free energy density~\eqref{eq:dfen} and discussed the additional information that can be extracted from it.
We derived a system of integral equations for the bulk part~\eqref{eq:bf} of the dynamical free energy density after a quench in the spin-$\frac{1}{2}$ XXZ chain. A qualitative analysis of the equations revealed a connection between singularities at large time and the absence of an energy  gap in the elementary excitations of the generalized Hamiltonian~\eqref{eq:genH}. 
We have shown that the non-analytic behavior is not peculiar to quenches across a critical point.

\begin{acknowledgements}
I thank Fabian Essler for lively discussions and illuminating remarks. I also thank Neil Robinson for useful comments. 
\end{acknowledgements}

\onecolumngrid
\newpage
\section{Supplemental Material}
In Ref.~\cite{ggeXXZ} it was considered the nonequilibrium time evolution \mbox{$\ket{\Psi(t)}=e^{-i H_{\rm XXZ}^{(\Delta)} t}\ket{\Psi_0}$}, with $\Delta=\cosh\eta>1$. Assuming that the state locally relaxes to a generalized Gibbs ensemble, Ref.~\cite{ggeXXZ} shown that  correlators in the stationary state can be computed with the same formalism developed for thermal correlators (see \cite{thermalcorr} and references therein), with the important difference that the stationary state is now determined by the local integrals of motion.
In particular, correlators can be expressed in terms of two $\pi$-periodic functions $\gothb$ and $\bar\gothb$ that satisfy the system of nonlinear integral equations
\be
\ba
{}\log\gothb(x)-\log\bar\gothb(x)=&h+[(k_++k)\ast\log(1+\gothb)](x)-[(k_-+k)\ast\log(1+\bar\gothb)](x)\ ,\\
g_\mu^+(x)=&-d(x-\mu)+\Bigl[k\ast\frac{g_\mu^+}{1+\gothb^{-1}}\Bigr](x)-\Bigl[k_-\ast\frac{g_\mu^-}{1+\bar{\gothb}^{-1}}\Bigr](x)\ ,\\
g_\mu^-(x)=&-d(x-\mu)+\Bigl[k\ast\frac{g_\mu^-}{1+\bar\gothb^{-1}}\Bigr](x)-\Bigl[k_+\ast\frac{g_\mu^+}{1+\gothb^{-1}}\Bigr](x)\ ,\\
4
k(\mu)+\frac{4 i}{\eta}\Omega_{\Psi_0}(-2\mu/\eta)=&-\int_{-\frac{\pi}{2}}^{\frac{\pi}{2}}\frac{\mathrm d
  x}{\pi}d(x)\Bigl(\frac{g^+_{\mu}(x)}{1+\gothb^{-1}(x)}+\frac{g^-_{\mu}(x)}{1+\bar\gothb^{-1}(x)}\Bigr)\ ,\\
4m^z=&\int_{-\frac{\pi}{2}}^{\frac{\pi}{2}}\frac{\mathrm d x}{\pi}\Bigl(\frac{g_{0}^{+}(x)}{1+\gothb^{-1}(x)}-\frac{g_{0}^{-}(x)}{1+\bar\gothb^{-1}(x)}\Bigr)\, ,
\ea
\ee
where \mbox{$k(x)=\sum_{n}\frac{e^{2i n x}}{e^{2 \eta |n|}+1}$}, \mbox{$k_\pm(x)=k(x\pm i\eta\mp i 0^+)$}, \mbox{$d(x)=\sum_{n}\frac{e^{2i n x}}{\cosh(\eta n)}$}, and \mbox{$[g_1\ast g_2](x)=\int_{-\pi/2}^{\pi/2}\frac{\mathrm d y}{\pi}g_1(x-y)g_2(y)$}; $m^z$ is the longitudinal magnetization $\frac{1}{2 L}\sum_\ell\braket{\Psi_0|\sigma_\ell^z|\Psi_0}$ and $\Omega_{\Psi_0}(x)$ is the generating function defined as
\be
\Omega_{\Psi_0}(x)=-i\sum_{k=1}\Bigl(\frac{\eta}{\sinh\eta}\Bigr)^k\frac{x^{k-1}}{(k-1)!}\frac{\braket{\Psi_0|H^{(k)}|\Psi_0}}{L}\, ,
\ee 
where $H^{(k)}$ are the minimal set of local conservation laws
\be\label{eq:LCL}
H^{(k)}=i\Bigl(\frac{\sinh \eta}{\eta}\frac{\partial}{\partial \lambda}\Bigr)^k\log\tau(i+\lambda)\Bigr|_{\lambda=0}\, .
\ee
Here $\tau$ is the XXZ transfer matrix, given by \mbox{$\tau(i+\lambda)=\mathrm{Tr}_{\rm aux}[\mathcal L_L(\lambda)\cdots \mathcal L_1(\lambda)]$}, where ``$\rm aux$'' denotes a auxiliary space, 
\be
\begin{gathered}
\mathcal L_j(\lambda)=\frac{1+\tau^z \sigma^z_j}{2}+A_\lambda\frac{1-\tau^z \sigma^z_j}{2}+B_\lambda(\tau^+\sigma^-_j+\tau^-\sigma^+_j)\\
A_\lambda=\frac{\sin(\frac{\eta\lambda}{2})}{\sinh(\frac{\eta\lambda}{2}+i\eta)}\, ,\qquad B_\lambda=\frac{i \sinh(\eta)}{\sinh(\frac{\eta\lambda}{2}+i\eta)}\, ,
\end{gathered}
\ee
and $\tau^\alpha$ are Pauli matrices acting on ``aux''.

If the longitudinal magnetization is zero and the initial state is parity invariant then $\Omega_{\Psi_0}(x)$ is even, $h=0$, and $\bar \gothb(x)=\gothb(-x)$; therefore everything can be expressed in terms of the single auxiliary function $\gothb(x)$.
The Lagrange multipliers are implicitly determined by Eq.~(14) of the main text, \emph{i.e.}
\be\label{eq:nlib}
\mathcal E[\gothb](x)=\frac{\sinh \eta}{2}\sum_{j=0}\lambda_{2j+1}\Bigl(\frac{\sinh\eta}{2}\frac{\partial}{\partial x}\Bigr)^{2j} d(x)\, ,
\ee
where the functional $\mathcal E[\gothb](x)$  is given by
\be
\mathcal E[\gothb](x)=[k\ast \log(1+\gothb)](x)-[k_+\ast\log(1+\gothb)](-x)-\log\gothb(x)\, .
\ee
The logarithm of the partition function of the generalized Gibbs ensemble $\rho_{\rm GGE}$ has the same form as in the thermal case and can be written as
\be\label{eq:F}
\lim_{L\rightarrow\infty }\frac{\log\mathrm{Tr}[e^{-\sum_i\lambda_i H^{(i)}}]}{L}\equiv \mathcal F[\gothb]=-i\int_{-\frac{\pi}{2}}^{\frac{\pi}{2}}\frac{\mathrm d x}{\pi}\frac{\sinh^2\eta\cot(x-i\frac{\eta}{2}) \log\bigl(1+\gothb(x)\bigr)}{\sin(x+i\frac{\eta}{2})\sin(x-i\frac{3\eta}{2})}-\int_{-\frac{\pi}{2}}^{\frac{\pi}{2}}\frac{\mathrm d x}{\pi}K_\eta(x) \log \gothb(x)\, ,
\ee
where $K_\eta(x)=\frac{\sinh\eta}{\cosh\eta-\cos(2x)}$.

As discussed in the main text, the bulk part of the dynamical free energy density con be computed by taking the expectation value of the time evolution operator on the GGE. The Loschmidt overlap in the GGE is the ratio of two partition functions 
\be
G_{\rm GGE}(t)\sim e^{-L f_{\rm bulk}(t)}\sim \frac{\mathrm{Tr}[e^{-\sum_i\lambda_i H^{(i)}}e^{- i H^{(1)} t}]}{\mathrm{Tr}[e^{-\sum_i\lambda_i H^{(i)}}]}
\ee
therefore (in the thermodynamic limit)  $f_{\rm bulk}(t)$ reads (\emph{cf.} \eqref{eq:F})
\be
f_{\rm bulk}(t)=\mathcal F[\gothb]-\lim_{L\rightarrow\infty}\frac{1}{L}\log\mathrm{Tr}[e^{-\sum_i\lambda_i H^{(i)}}e^{- i H^{(1)}t}]\, .
\ee
The second term is the logarithm of the partition function of the generalized model in which the Lagrange multiplier of the Hamiltonian is shifted by $i t$.
Because this modification does not break parity symmetry, the modified partition function can be expressed in terms of a auxiliary function $\gothc_t(x)$ that satisfies the same equations of $\gothb(x)$ \eqref{eq:nlib} but with the shifted Lagrange multiplier. 
After some algebra we obtain Eq.~(18) of the main text.

\begin{figure}
\includegraphics[width=0.75\textwidth]{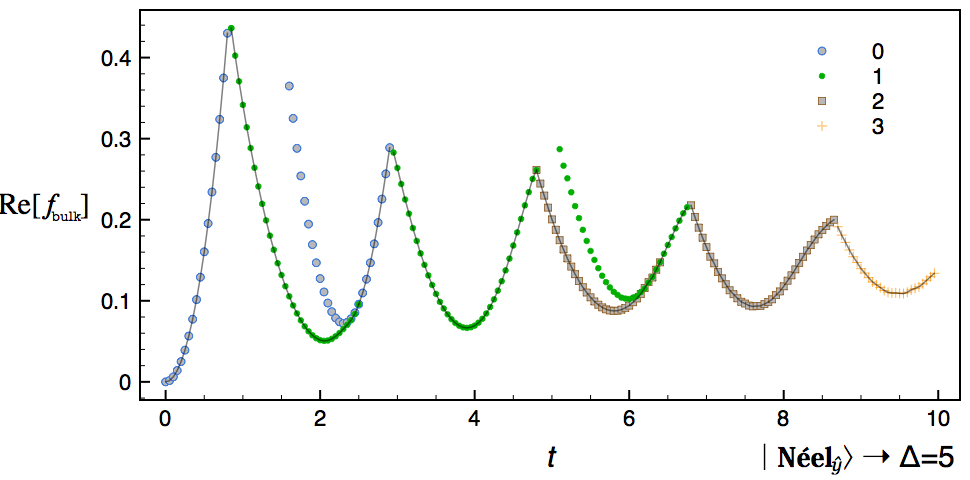}
\caption{The real part of  $f_{\rm bulk}(t)$ for the nonequilibrium time evolution with the XXZ Hamiltonian with $\Delta=5$ starting from the N\'eel state $\ket{\cdots\uparrow\downarrow\uparrow\downarrow\cdots}$ with spins aligned along the transverse direction $\hat y$. We assumed that equations (18) and (19) of the main text can be applied also when $f(t)$ is not regular, as in the present case. In the legend it is reported the winding number of $1+\gothc_t(x)$. There are two time intervals in which we find two solutions with different winding number. We have followed the prescription of imposing continuity going forward in time. The two solutions seem to join with the same first derivative.}
\label{fig:2}
\end{figure}

We notice that this simple derivation might break down if the curve (in the complex plane) $1+\gothc_t(x)$, for $-\frac{\pi}{2}<x\leq \frac{\pi}{2}$, has nonzero winding number about the origin. Figure~\ref{fig:2} is an attempt to find a continuous solution for a quench with dynamical phase transitions. We have not investigated whether continuity  is the correct prescription.  
We however stress that in Figure~\ref{fig:2} there are times at which there is a different kind of transition, where it seems that also the first derivative of the dynamical free energy density is  continuous. It must be clarified whether this can be interpreted as a ``higher order dynamical phase transition'' or it is an artifact of having used formulae beyond their regime of validity.  

A more detailed analysis will be reported in a future work. 

\end{document}